\documentclass[twocolumn,showpacs,floatfix,superscriptaddress,amsmath,amssymb,pre]{revtex4}
\usepackage{mathrsfs}
\usepackage{txfonts}
\usepackage{amssymb}
\usepackage{graphicx}
\usepackage{hyperref}
\usepackage{ulem}
 \usepackage{overpic}
\usepackage{color}
 \usepackage{psfrag}
\usepackage{tabularx}
\usepackage{array}
\usepackage{placeins}
\makeatletter

\begin{document}

\title{Quantum entanglement entropy and Tomonaga-Luttinger liquid to liquid transition in biquadratic spin-1 XY chain with rhombic single-ion anisotropy}

\author{Yan-Wei Dai}
\affiliation{Centre for Modern Physics and Department of Physics,
Chongqing University, Chongqing 400044 China}

\author{Yao Heng Su}
\affiliation{School of Science, Xi'an Polytechnic University, Xi'an 710048 China.}

\author{Sam Young Cho}
\altaffiliation{E-mail: sycho@cqu.edu.cn}
\affiliation{Centre for Modern Physics and Department of Physics,
Chongqing University, Chongqing 400044 China}

\author{Huan-Qiang Zhou}
\affiliation{Centre for Modern Physics and Department of Physics,
Chongqing University, Chongqing 400044 China}

\begin{abstract}
 Quantum phase transitions are investigated in biquadratic spin-$1$ XY chain with rhombic single-ion anisotropy by using the ground state energy, the bipartite entanglement entropy, and the mutual information. It turns out that there are three spin nematic phases and two Tomonaga-Luttinger (TL) liquid  phases with the central charge $c = 1$ for the whole parameter space. The TL Liquid phases emerge roughly for biquadratic interaction strength two times stronger than the absolute value of the single-ion anisotropy. The ground state energy and the derivatives up to the second order reveal a first-order quantum phase transition between spin nematic ferroquarupole (FQ) phases but cannot capture an evident signal of transitions between the spin nematic phases and the TL Liquid phases as well as transition between the two TL liquid phases.  The TL liquid-to-liquid transition point features a highly degenerate state and the spin-block entanglement entropy increases logarithmically with block size. The bipartite entanglement entropy exhibits a divergent or convergent behavior identifying the TL Liquid or spin nematic FQ phases, respectively. Similarly,  The mutual information and the spin-spin correlation are shown to decay algebraically or exponentially with increasing the lattice distance in the TL Liquid or spin nematic FQ phases, respectively. In the TL liquid phase, the exponents $\eta_I$ and $\eta_z$ of the mutual information and the spin-spin correlation vary with the interaction parameter of the biquadratic interaction strength and the rhombic single-ion anisotropy and satisfy the relationship of $\eta_z <\eta_I$. Such changes of characteristic behavior of the bipartite entanglement entropy, the mutual information and the spin-spin correlation indicate an occurrence of the Berezinskii-Kosterlitz-Thouless (BKT)-type quantum phase transition between the TL Liquid phase and the spin nematic FQ phase. The staggered spin fluctuation $\langle S^x S^y \rangle$ is shown to play a significant role for the emergence of the TL liquid phase and thus give rise to the BKT-type quantum phase transition.
\end{abstract}


\maketitle
\section{Introduction}
 Quantum fluctuations \cite{Hertz,Vojta,Campisi} attributed to the uncertainty principle affect matter more and more strongly as temperature becomes very low and cause quantum phase changes in matter \cite{Sachedev99,Sachedev23}. These quantum phase transitions and quantum critical phenomena \cite{Hertz,Vojta,Sachedev99,Sachedev23,Sachdev,Wen,Carr,Ashida} are at the heart of universal low-energy properties in quantum many-body systems and are crucial for understanding the fundamental physics of condensed matter physics. A boundary between two macroscopically distinguishable phases is known to be related to singularities in the derivatives of the free energy \cite{Reichl,Wu}. Such a phase transition, corresponding to a singular behavior of ground state energy, is usually described by the Landau-Ginzburg-Wilson (LWG)'s paradigm of spontaneous symmetry breaking \cite{Landau58} and as such is detectable by a corresponding local order parameter being nonzero value in broken-symmetry phase \cite{Anderson,Coleman}.

 Further, the discovery of the quantum Hall effect ~\cite{Klit}, not being understood well by local order parameters within the mechanism of LWG symmetry breaking, has encouraged to establish an existence of quantum phases and quantum phase transitions beyond the LGW paradigm \cite{Wen}. In other words, without breaking a (explicit) symmetry, quantum phase of matter can be well characterized as a non-local (topological) order \cite{Tsui,Wen5} rather than a local order and quantum phase transitions, i.e., so-called topological quantum phase transition \cite{Wen1,Kitaev03,Trebst,Hamma,Cast,Feng,Chen,Yu,JVidal2}, can occur between these two different phases. Examples on studying topological orders and phase transitions include fractional quantum Hall effects~\cite{Tsui,Laughlin}, Haldane phase~\cite{Haldane,Wang13}, chiral spin liquids~\cite{Kalmeyer,Wen6}, and $Z_2$ spin liquids~\cite{Read,Wen7,Moessner}. Another examples are to be quantum crossovers \cite{Khan,Mao}, being an adiabatic connection between two orthogonal states without an explicit phase transition, and infinite-order quantum phase transitions \cite{J.Zhang}, which cannot be identified due to the inherent limitation of detecting nonanalyticities of the ground state energy and the finite-order derivatives. For instance, quantum crossovers are studied in the Bardeen-Cooper-Schrieffer (BCS)- Bose-Einstein condensate (BEC) crossover \cite{Khan,Leggett,Nozieres,Perali} and in the biaxial spin nematic to nematic crossover \cite{Mao}. Infinite-order quantum phase transitions can occur for gapped-to-gapped Gaussian phase transitions \cite{Ovchinnikov,Haldane1982,Nakamura,Chen2003,Zhang2021} and  gapped-to-gapless Berezinskii-Kosterlitz-Thouless (BKT) phase transitions \cite{Berezinskii,Berezinskii71,Kosterlitz72,Kosterlitz73,Kosterlitz74,Hu20}.

 Along with the great progresses in understanding such quantum phases and quantum phase transitions in quantum many-body systems, novel quantum spin states and quantum phase transitions are being found in various quantum spin systems \cite{Book_Schollwock,Lacroix,Diep,Udagawa,Mao}. To be specific, one of the most fascinating spin states is spin nematic states  \cite{Blume1969,Matveev1973,Papanicolaou1988,Tsunetsugu2006,Lauchli2006,Penc2011,Andreev1984,Sudan,Sakai2022,Zhitomirsky,Balents,Ueda} that have no magnetic order, i.e., no long-range magnetic order. In spin nematic states as a nonmagnetic state, spin fluctuations become thus more important and aligned spin fluctuations can characterize the states. Aligned on-site spin quadrupole or higher-rank multipolar moments \cite{Sudan,Sakai2022}, for instance, induced from anisotropic spin fluctuations can form an orientational order, as a simplest case, which can break the spin rotation symmetry without magnetic (dipole moment) order.  Quantum spin fluctuations can induce other types of distinct nematic phases without magnetic order \cite{Lauchli2005,Shannon2006,Sindzingre}. Conventional experiments are known to be difficult to detect even the simplest cases \cite{Penc2011,Smerald2015,Smerald2016,Kohama2019}. Remarkably, such a quantum spin nematic phase has been observed recently for the first time through dipole-quadrupole interference in circular dichroic resonant x-ray diffraction in a square-lattice iridate \cite{Kim2024}.

 Identifying and characterizing different types of spin nematic phases and quantum phase transitions among them beyond the BGW mechanism becomes a significant and interesting issue for a deeper understanding of the exotic quantum phenomena of quantum spin systems. Meanwhile, various quantum measures such as quantum entanglement \cite{Wootters,vidaletal,RMP2005,SYC06,Amico2008,Eisert,Modi,Preskill,Osborne2002,Vidal1,Evenbly,Kitaev,Tagliacozzo,Pollmann}, quantum mutual information \cite{Wolf2008,Groisman2005,Adami1997,Alcaraz2015,Schumacher2006,Dai2018,Dai2021} and quantum coherence measures \cite{Baumgratz,JJChen,Karpat,QChen,Yi,Thakur,Sha,Li,Hu,Malvezzi,Chao}, introduced to measure quantumness of physical systems from the perspective of information theory, have been shown to be very useful tools in exploring the states of many-body systems. Even if it is not known a priori whether there is a quantum phase transition or what kind of phase there is at all for a given many-body system, such quantum measures have been shown to able to  explore quantum coherence or entanglement aspect of fundamental feature of quantum phases and quantum phase transition as well as identify quantum phase transitions. Thus, in our study, we investigate quantum entanglement, quantum mutual information and quantum quadrupole moments to identify and classify quantum critical phenomena with spin nematic phases and quantum phase transitions between critical phases or between critical and spin nematic phases.

 One-dimensional infinite spin-$1$ lattice with biquadratic XY interaction and rhombic single-ion anisotropy is introduced to investigate spin nematic phases and quantum critical phenomena as well as quantum phase transitions between them. The infinite matrix product state (iMPS) representation \cite{Vidal2003,Latorre,Vidal07,Tagliacozzo,Pollmann} is employed and the infinite time-evolving block decimation (iTEBD) algorithm \cite{Vidal07,Tagliacozzo,Pollmann,Su12,Su13} is used to calculate the ground state wave functions. Based on the iMPS ground states, as the very conventional way to clarify quantum phase transitions, the ground state energy and the derivatives up to the second order are investigated and disclose only a first-order discontinuous quantum phase transition between spin nematic ferroquadropole (FQ) phases.
 However, the characteristic features of the bipartite entanglement entropy separate gapless critical phases with the central charge $c = 1$ and gapped noncritical nematic phases, respectively, and further the mutual information and the spin-spin correlation show the features of Tomonaga-Luttinger (TL) liquid \cite{Tomonaga,Luttinger,Haldane81,Belavin,Cardy84,Affleck86,Tsvelik,Lake,Zaliznyak,Giamarchi,Takahashi} state in the critical phases, which indicates an occurrence of a BKT-type quantum phase transition \cite{Berezinskii,Berezinskii71,Kosterlitz72,Kosterlitz73,Kosterlitz74} which cannot be identified in finite derivative of the ground state energy. Furthermore, we find that a continuous quantum phase transition occurs between the two TL liquid phases and the spin chain system has a high degenerate ground state at the TL liquid to liquid transition point. The characteristics of the diagonal spin quadrupole moments classify the spin nematic FQ phase, while the spin fluctuation $\langle S^x S^y \rangle$ with the spin-$1$ operators $S^{x/y}$ acts as an important fator for the emergence of the TL liquid phase and the occurrence of the BKT-type quantum phase transition.

 This paper is organized as follows. In Sec. \ref{section2}, the infinite biquadratic spin-$1$ XY chain with the rhombic-type single-ion anisotropy is introduced. The iMPS approach is briefly explained and the iTEBD algorithm is employed in calculating ground state wave functions. Section \ref{section3} focuses on studying the nonanalyticity of ground state energy and the derivatives up to the second order. Section \ref{section4} discusses the characteristic divergent or convergent behaviors of the bipartite entanglement entropy with increasing the truncation dimension to identify the massless (TL Liquid) or massive phases, respectively. We also discuss the singularity of the bipartite entanglement entropy representing the quantum phase transitions including the BKT-type quantum phase transitions and the TL liquid to liquid transition. The characteristics of the ground state is studied at the TL liquid to liquid transition point. In Sec. \ref{section5}, the mutual information, being the sum of classical and quantum correlations, is discussed in characterizing the massless and massive phases. For comparison with the mutual information, the spin-spin correlation is also discussed on an equal footing. In Sec. \ref{section6}, we find that the diagonal components of the quadrupole moment tensor identify the spin nematic FQ phases and the off-diagonal components seem to play a significant role for the TL Liquid phases. A summary and remarks of this work is given in Sec. \ref{summary}.

\section {Biquadratic spin-$1$ XY model with rhombic single-ion anisotropy}
\label{section2}
 Let us consider an one-dimensional infinite spin-$1$ chain with biquadratic spin-$1$ XY interaction with rhombic single-ion anisotropy. The system Hamiltonian can be written as
\begin{equation}
 H = \sum_{i=-\infty}^\infty \cos\theta(S_{i}^{x}S_{i+1}^{x}+S_{i}^{y}S_{i+1}^{y})^2
   + \sin\theta [(S_{i}^{x})^2-(S_{i}^{y})^2],
\label{ham1}
\end{equation}
 where the biquadratic exchange interaction and the rhombic single-ion anisotropy are varied with $\cos\theta$ and $\sin\theta$, respectively, in the circular parameter space $\theta$.
 $S^{\alpha}_i (\alpha \in \{x,y,z\})$ is the spin-$1$ operator at $i$ site.
 In fact, the \textit{ferromagnetic} biquadratic spin-$1$ XY model with the rhombic single-ion anisotropy was introduced to study the spin nematic to nematic transition and to answer on how and to what extent quantum phase transition can be understood only by interpreting the behaviors of quantum coherence measures ~\cite{Mao}. In this study, we focuses on mainly quantum entanglement, mutual information, and quantum quadrupole moments to identify gapless critical (TL liquid) phases and capture the quantum phase transitions such as a TL liquid to spin nematic phase transition and a TL liquid to liquid phase transition.

   \begin{figure}
    \includegraphics[width=0.4\textwidth]{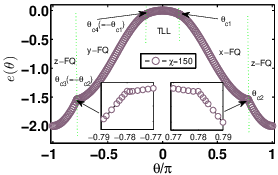}
    \caption{(color online)  Ground state energy per site $e(\theta)$ as a function of
    $\theta$ for the truncation dimension $\chi=150$. The insets shows that the nonanalyticities of the ground state energy appear in the form of the kinks (i.e., at $\theta =\pm\theta_{c2}$).
    } \label{Fig1}
     \end{figure}
 In order to calculate the ground state of our model Hamiltonian (\ref{ham1}),
 the iMPS representation in one-dimensional infinite spin lattices is employed for numerical
 study with a wave function $\left|\psi\right\rangle$ of the Hamiltonian (\ref{ham1}).
 The iTEBD algorithm \cite{Vidal07, Tagliacozzo, Pollmann, Su12, Su13} can bring the initial wave function $\left|\psi\right\rangle$ into the numerical ground stat $\left|\psi_G\right\rangle$, the iTEBD procedure \cite{Vidal07} has been completed with the time step decreased from $dt = 0.1$ to $dt = 10^{-6}$ according to a power law when the chosen initial state approaches to a ground state. During the procedure, the convergence of energy yields a ground state wavefunction $|\psi_G\rangle$ in the iMPS representation for a given truncation dimension. Any reduced density matrix can be given by using the full ground state density matrix $\varrho_G = |\psi_G\rangle\langle\psi_G|$. For example, the reduced density matrix $\varrho_{AB}$ of two spins $A$ and $B$ is given by tracing out the degrees of freedom of the rest of the two spins, i.e., $\varrho_{AB} = \mathrm{Tr}_{(AB)^c} \, \varrho_G$ with the full description of the ground state $|\psi_G\rangle$ in a pure state. Necessitate reduced density matrices are given from the ground state wave function for this study.

\section{Ground state energy and quantum phase transitions}
\label{section3}
 At zero temperature, a strong quantum fluctuation can induce a structural change of many-body ground state wave function drastically. Such a change of ground state wave function can bring quantum many-body systems into a different phase of matter. In the conventional framework, ground state energy of quantum many-body systems features its intrinsic nonanalyticities, for instance, such as the energy level crossings, signaling corresponding quantum phase transitions. Searching nonanalyticities of ground state energy is the very conventional classification, e.g., a finite discontinuity or divergence of $n$th-order derivative of ground state energy characterizes a $n$th-order phase transition \cite{Reichl, Wu}. In this section, we discuss the ground-state energy per site $e$ in order to study quantum phases and phase transitions systematically, i.e., to notice which order quantum phase transition can occur with the competition of the biquadratic interaction and the rhombic single-ion anisotropy in the spin-$1$ chain Hamiltonian (\ref{ham1}).

 In Fig. \ref{Fig1}, we plot the iMPS ground state energy per site $e(\theta)$ as a function of the angle ratio $\theta$ of the biquadratic interaction and the rhombic anisotropy for the truncation dimension $\chi=150$. The ground state energy is continuous in the whole parameter space $\theta$. One can notice the two kinks (actually, at $\theta=\pm \theta_{c2}=\pm 0.778\pi$), as shown clearer in the insets of Fig. \ref{Fig1}. The kinks indicate nonanalyticities of the ground state energy, which implies that an energy level crossing occurs at each kink. Since a level crossing gives rise to a finite discontinuity of first-order derivative of ground state energy, each kink indicates an occurrence of first-order quantum phase transition. Except for those kinks, no other nonanalyticity is noticeable in the ground state energy, as shown in Fig. \ref{Fig1}. Accordingly, the ground state energy in Fig. \ref{Fig1} shows a possibility occurring first-order quantum phase transitions at the kinks.

   \begin{figure}
    \includegraphics[width=0.4\textwidth]{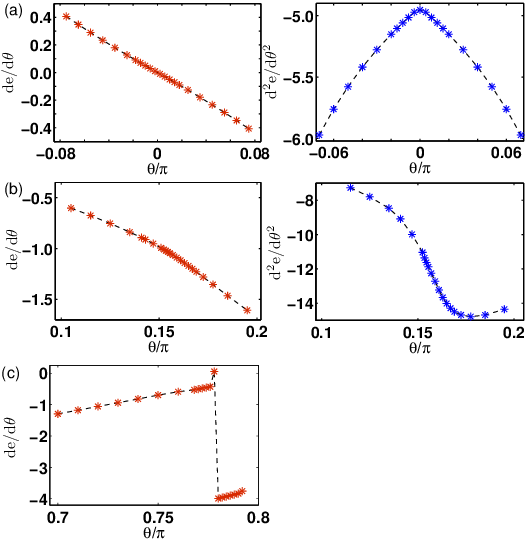}
    \caption{(color online) First- and second-derivatives of ground state energy $e(\theta)$ as a function of $\theta$ near (a) $\theta =\theta_{c0} = 0$, (b) $\theta = \theta_{c1}$, and (c) $\theta=\theta_{c2}$.
    } \label{Fig2}
     \end{figure}

 A simple classification of quantum phase transition is to perform numerical derivatives of the ground state energy $e(\theta)$ of Fig. \ref{Fig1} and classify phase transitions according to the nonanalyticity of the derivatives \cite{Reichl}. We have investigated the numerical derivatives of the iMPS ground state energy up to the second-order derivatives and the result can be summarized in three points, as shown in Fig. \ref{Fig2}. Figure \ref{Fig2} (a) shows the derivatives of the ground state energy near $\theta=0$. The first- and the second-order derivatives are continuous. One can notice a upward-facing cusp at $\theta = 0$ in the second derivative and thus expect a third-order quantum phase transition denoted by a finite discontinuity of the third-order derivative at $\theta = \theta_{c0} = 0$. However, numerically reaching to a reliable higher or much higher derivatives is to be a very difficult task. In fact, it can be difficult to reach reliable precision of higher-order numerical derivatives in our calculations, which is beyond the scope of our work. As a result, it is not possible to determine whether the finite discontinuity or divergence of third derivative appears at $\theta=0$. Thus, we cannot get an answer as to whether the quantum phase transition occurs at $\theta=0$ due to the limitation of the numerical calculation.

 Whereas Fig. \ref{Fig2}(b) shows that the first- and second--order derivatives of the ground state energy are continuous and exhibit no nonanalyticity for the chosen parameter region (actually, near $\theta = \theta_{c1}$). In fact, except for $\theta = \theta_{c2}$ in Fig. \ref{Fig2} (c), the first- and second-order derivatives of the ground state energy exhibit their continuity and no singular behavior. Obviously, the expected first-order phase transition at the kink $\theta = \theta_{c2}$ in Fig. \ref{Fig1} is confirmed in the finite discontinuity of the first-order derivative at $\theta = \theta_{c2}$, as shown in Fig. \ref{Fig2}(c). Up to the numerical second-order derivatives, as a result, the nonanalyticities of ground state energy disclose clearly only the occurrence of the first-order quantum phase transitions at $\theta = \pm\ \theta_{c2}$.

 Due to the limitation of our numerical calculation, we have studied the possibility of quantum phase transition up to the second-order derivatives of the ground state energy in the biquadratic spin-$1$ XY chain with the rhombic single-ion anisotropy according to the classical definition of the corresponding phase transitions given in terms of the free energy \cite{Reichl,Wu}. However, if another phase transition occurs at other parameter $\theta$ of the Hamiltonian ($\ref{ham1}$), it means that a third- or higher-order phase transition occurs, including an infinite-order quantum phase transition such as a BKT-type quantum phase transition \cite{Berezinskii,Berezinskii71,Kosterlitz72,Kosterlitz73,Kosterlitz74}. Of course, if another phase transition does not occur, it cannot be excluded quantum crossover \cite{Mao,Khan}, i.e., an adiabatic connection of two orthogonal states without explicit phase transition at a specific parameter such as the BCS-BEC crossover \cite{Leggett,Nozieres,Perali}. Hence, the bipartite entanglement entropy will be considered to further capture and  classify distinct phase transitions in our spin chain in the next section, since the bipartite entanglement entropy can capture characteristic phase transitions without a priori knowledge of any kind of phase transitions for a given many-body system \cite{vidaletal,RMP2005,SYC06,Amico2008,Eisert,Modi,Preskill,Osborne2002,Vidal1,Evenbly,Kitaev,Tagliacozzo,Pollmann}.

\section {Entanglement entropy and quantum phase transitions}
\label{section4}
 For a given system, roughly, if two subsystems are not independent each other, they can be said entangled. Quantum mechanically, for such a entangled state, performing a local measure instantaneously affect the outcome of local measurements far away. Entanglement between subsystems is then one of the most fundamental and fascinating features of quantum mechanics. Various measures of entanglement have been suggested and demonstrated to be a useful tool for detecting and classifying quantum phase transitions~\cite{vidaletal,RMP2005,SYC06,Amico2008,Eisert,Modi,Preskill,Osborne2002,Vidal1,Evenbly,Kitaev,Tagliacozzo,Pollmann}.
 For quantum spin chains, as a bipartite entanglement measure, the von Neumann entropy is shown to exhibit qualitatively different behaviors at and off criticality~\cite{Vidal2003,Calabrese,Korepin,CC09}. For instance, the entanglement entropy of a subsystem formed by a block of contiguous $n$ sites of an infinite system has the leading behavior $S = \frac{c}{3} \log_2 n$ if the system is critical, or $S = \frac{c}{3} \log_2 \xi$ if the system is near critical, with correlation length $\xi$ and central charge $c$ of the conformal field theory describing the critical point~\cite{Calabrese,Korepin,CC09}. Accordingly, the bipartite entanglement can capture a quantum phase transition. In order to confirm the results of quantum phase transitions from the ground state energy and further capture distinct quantum phase transitions, we will thus consider the bipartite entanglement entropy between the two semi-infinite chains split in the infinite spin chain.

 In the iMPS approach, an iMPS ground state can be expressed in terms of the left and right bases $|\psi_i^L \rangle$ and $|\psi_i^R \rangle$, where $|\psi_i^L \rangle$ and $|\psi_i^R \rangle$ are the Schmidt bases for the left and right semi-infinite spin chains, respectively. The iMPS ground state is written as $|\psi_G\rangle = \sum_i^\chi \lambda_i |\psi^L_i\rangle |\psi^R_i\rangle$ with the Schmidt coefficients $\lambda_i$ and the truncation dimension $\chi$. The reduced density matrices for the left and right semi-infinite chains can be obtained from the full ground state density matrix $\varrho_G = |\psi_G \rangle \langle \psi_G|$, i.e., tracing out the degrees of freedom of the rest of the two semi-infinite chains as $\varrho_{L/R} = \mathrm{Tr}_{R^c/L^c} [ \varrho_G ]$. The von Neumann entropy for the bipartite entanglement is given as $S = -\mathrm{Tr} \varrho_L \log_2 \varrho_L = - \mathrm{Tr} \varrho_R \log_2 \varrho_R$. In terms of the Schmidt coefficients $\lambda_i$, the von Neumann entropy is given as ~\cite{Tagliacozzo, Pollmann}
\begin{equation}
   S =-\sum_{i=1}^{\chi}\lambda_i^2\log_2\lambda_i^2.
   \label{Entropy}
\end{equation}

\begin{figure}
\includegraphics[width=0.4\textwidth]{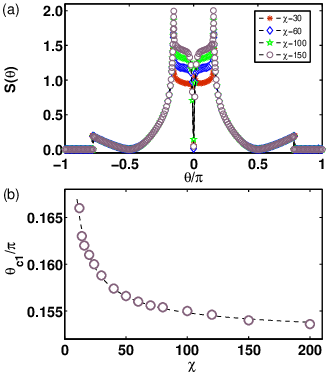}
 \caption{(color online) (a)The bipartite entanglement entropy $S(\theta)$ as a function of $\theta$, with the truncation dimension $\chi=30$, $60$, $100$ and $150$, respectively. The five phase transition points are located from the entanglement entropy $S(\theta)$. (b) An extrapolation with respect to the truncation dimension $\chi$ is performed for the pseudo critical points $\theta_{c1} (\chi)$ with the fitting function being $\theta_{c1}(\chi)=a\chi^b + \theta_{c1}(\infty)$, the critical point $\theta_{c1}(\infty)$ is extracted in the thermodynamic limit $\chi \rightarrow \infty$.}
  \label{Fig3}
  \end{figure}
 \subsection{Transition points and bipartite entanglement entropy }
 We have calculated the iMPS ground states for various truncation dimensions $\chi = 30$, $60$, $100$ and $150$. Figure \ref{Fig3} (a) shows the bipartite entanglement entropy $S(\theta)$ as a function of $\theta$. One can first see the discontinuities of abrupt jumps that stand out in the bipartite entanglement entropy. The discontinuities correspond to the first-order quantum phase transitions at $\theta = \pm \theta_{c2}$, as was shown in Figs. \ref{Fig1} and \ref{Fig2} (c). Together with the discontinuous quantum phase transition at $\theta =\pm \theta_{c2}$, one can notice more three singular points of the bipartite entanglement entropy, i.e., one downward-facing cusp at $\theta=0$ and two upward-facing cusps at $\theta=\pm\, \theta_{c1}$. The three noticeable downward- and upward-facing cusps of the bipartite entanglement entropy at $\theta=\theta_{c0}=0$ and $\theta=\pm \theta_{c1}$, indicating an occurrence of continuous quantum phase transitions, do not appear to have any correspondence to nonanalyticities of the ground state energy in Figs. \ref{Fig1} and \ref{Fig2} because the ground state energy seems to be analytic up to the second-order derivatives as shown in Fig. \ref{Fig2} (b). A consistent interpretation for both the singular behaviors of the  ground state energy and the bipartite entanglement entropy would be that there occurs a higher-order continuous quantum phase transition, i.e., from third- to infinite-order phase transitions, any one of these quantum phase transitions can occur.

 Some of the five singularities move and some rarely move, as the truncation dimension $\chi$ increases. One can notice from Fig. \ref{Fig3} (a) clearly that $\theta_{c1}$ shifts but $\theta_{c0}$ and $\theta_{c2}$ do not. In the thermodynamic limit $\chi \rightarrow \infty$, the critical points can be estimated by the extrapolation with the numerical fitting function $\theta_{c1}(\chi)=a\, \chi^b + \theta_{c1}(\infty)$ and the fitting constants $a$ and $b$ \cite{Tagliacozzo}. The cusps at $\theta=\theta_{c1}$ and $\theta = -\theta_{c1}$ move closer on the parameter space $\theta$ as the truncation dimension $\chi$ increases from $\chi=30$ to $\chi=150$. In Fig.~\ref{Fig3} (b), we plot the transition points $\theta_{c1}(\chi)$ as a function of the truncation dimension $\chi$ and perform the extrapolation in order to obtain the critical point $\theta_{c1}(\infty)$. The best fits gives the critical point $\theta_{c1}(\infty)=0.152(1) \pi$ with $a=0.09(4)\pi$ and $b=-0.8(2)$.

\subsection{Gapless critical (Tomonaga-Luttinger liquid) and gapped noncritical massive phases}
\label{Luttinger}
\begin{figure}
\includegraphics[width=0.4\textwidth]{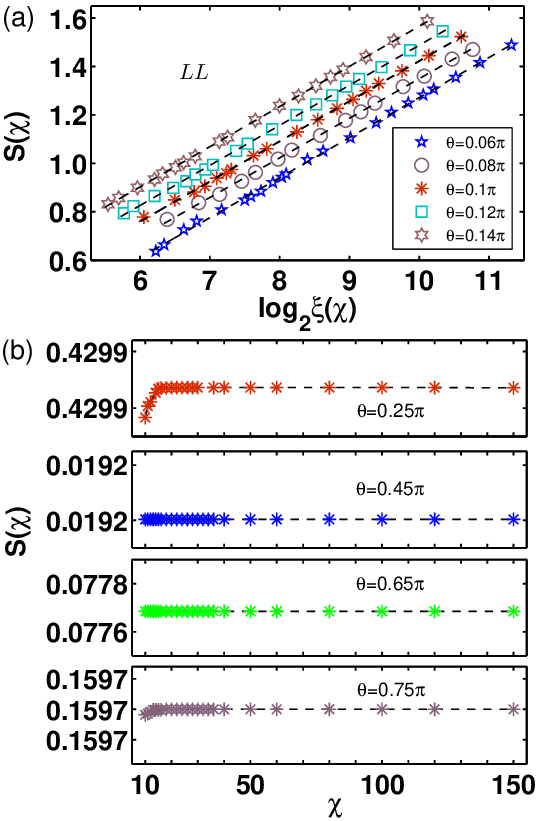}
 \caption{(color online) (a) Bipartite entanglement entropy $S(\chi)$ versus $\log_2\xi(\chi)$ at five points $\theta=0.06\pi$, $0.08\pi$, $0.1\pi$, $0.12\pi$, and $0.14\pi$ in the gapless Tomonaga-Luttinger liquid phases with the bond dimension ranging from $\chi=16$ to $\chi=200$. The black dotted lines denote the fitted line. (b) Bipartite entanglement entropy $S(\chi)$ as a function of the truncation dimension $\chi$ at $\theta =0.25\pi$, $0.45\pi$, $0.65\pi$, and $0.75\pi$ in the gapped nonmagnetic phases, i.e., ferroquarupole phases.}
  \label{Fig4}
  \end{figure}

 One can easily notice a change of the amplitude of the bipartite entanglement entropy for $-\ \theta_{c1}(\infty) < \theta < 0$ and $0 < \theta < \theta_{c1}(\infty)$ as the truncation dimension varies from $\chi=30$ to $\chi=150$ in Fig. \ref{Fig3} (a). To be specific, for $-\ \theta_{c1}(\infty) < \theta < 0$ and $0 < \theta < \theta_{c1}(\infty)$, the bipartite entanglement entropy increases and can diverge with increasing the truncation dimension. Such divergences of the bipartite entanglement entropy are shown in Fig. \ref{Fig4}. Otherwise, the bipartite entanglement entropy does not change nearly and is to be saturated with increasing the truncation dimension. Actually, diverging behavior of the bipartite entanglement entropy implies that the system is in a massless (gapless critical) phase and also the divergence of the bipartite entanglement entropy can characterize universality classes of the massless phase through the central charge $c$ of the underlying conformal field theory \cite{Ginsparg,Srednicki,Callan,Fiola,Holzhey,Vidal2003,Latorre,Vidal07,Tagliacozzo,Pollmann}. Accordingly, the scaling behavior of the bipartite entanglement entropy then enables to distinguish whether the system is in either massless phases or phase with a mass gap \cite{Vidal2003,Latorre,Vidal07,Tagliacozzo,Pollmann}.

 In the iMPS representation, the central charge $c$ for $-\ \theta_{c1}(\infty) < \theta < 0$ and $0 < \theta < \theta_{c1}(\infty)$ can thus be studied and estimated from the scaling relations. Extracting the central charge from the scaling relation of bipartite entanglement entropy versus the correlation length is a very useful method to investigate the quantum many-body critical systems particularly in one-dimensional quantum spin models. A finite entanglement scaling analysis can be defined~\cite{Tagliacozzo, Pollmann}
\begin{equation}
   S(\chi) = \frac{c}{6} \log_2 \xi(\chi)+S_0,
 \label{ss}
\end{equation}
 where $\xi(\chi)\propto\chi^\kappa$ is the correlation length with $\kappa$ being the finite entanglement scaling exponent and $S_0$ is a fitting constant. In the iMPS representation, for a given truncation dimension $\chi$, the correlation length $\xi$ is characterized by the ratio of the largest $\varepsilon_1(\chi)$ to the second largest $\varepsilon_2(\chi)$  eigenvalues of the transfer matrix, i.e.,
 $\xi(\chi)=1/\log_2|\varepsilon_1(\chi)/\varepsilon_2(\chi)|$.

\subsubsection{Critical entanglement - Gapless Tomonaga-Luttinger liquid (TLL) phase (massless phase)}
\label{TLL}
 We have calculated the bipartite entanglement entropy for $-\ \theta_{c1}(\infty) < \theta < 0$ and $0 < \theta < \theta_{c1}(\infty)$. It turns out a symmetric behavior of the bipartite entanglement entropy, i.e., $S(\theta) = S(-\theta)$ for the chosen parameter points. We plot the bipartite entanglement entropy $S(\chi)$ as a function of $\log_2 \xi(\chi)$ at the five different points $\theta=0.06\pi$, $0.08\pi$, $0.1\pi$, $0.12\pi$, and $0.14\pi$ with the truncation dimension ranging from $\chi=16$ to $\chi=200$ in Fig. \ref{Fig4} (a). It is shown clearly that the bipartite entanglement entropy diverges logarithmically with the correlation length $\xi$, i.e., the system is in a gapless critical phase \cite{Vidal2003}. Thus, the finite entanglement scaling analysis \cite{Tagliacozzo, Pollmann} is performed and the fitted lines are denoted by the black dotted line in Fig.~\ref{Fig4} (a). The central charges are estimated by the best fits and listed in Table ~\ref{table1}. All the estimate central charges $c$ are very close to $1$ within a relative error being less than $2\%$. Accordingly, the gapless one-dimensional system with the central charge $c \simeq 1$ is in a gapless TL liquid phase \cite{Tomonaga,Luttinger,Haldane81,Belavin,Cardy84,Affleck86,Tsvelik,Lake,Zaliznyak} for $-\ \theta_{c1}(\infty) < \theta < 0$ and $0 < \theta < \theta_{c1}(\infty)$. As a characteristic of the gapless TL liquid phase, a power-law behavior of correlation functions will be discussed in the next Sec. \ref{section5}.

\begin{table}
\renewcommand\arraystretch{2}
\caption{Estimates for the central charge $c$ in the Tomonaga-Luttinger liquid phases
 obtained from the bipartite entanglement entropy.}
\begin{tabular}{ccccccc}
\hline\hline
      \begin{minipage}{1cm} $\theta$ \end{minipage}
      &\begin{minipage}{1cm}$0.06\pi$ \end{minipage} &
      \begin{minipage}{1cm} $0.08\pi$ \end{minipage} &
      \begin{minipage}{1cm} $0.1\pi$ \end{minipage} &
      \begin{minipage}{1cm} $0.12\pi$ \end{minipage} &
      \begin{minipage}{1cm} $0.14\pi$ \end{minipage}  \\
\hline
 \begin{minipage}{1cm}$c$  \end{minipage}
 & 0.99(2) & 0.981(1) & 0.99(1) & 0.99(1) &0.995(6) \\
\hline\hline
\end{tabular}
\label{table1}
\end{table}

\subsubsection{Noncritical entanglement - Gapped massive phases}
 In contrast to the TL liquid phases for $-\ \theta_{c1}(\infty) < \theta < 0$ and $0 < \theta < \theta_{c1}(\infty)$, Fig. \ref{Fig4} (b) shows that the bipartite entanglement entropy exhibits a simple saturation behavior at $\theta = 0.25 \pi$, $0.45 \pi$, $0.65\pi$ and $0.75\pi$ as the truncation dimension $\chi$ increases. Such saturation behaviors in the bipartite entanglement entropy indicate that the biquadratic spin-$1$ XY chain system with the rhombic anisotropy is in a noncritical ground state, i.e., a massive phase \cite{Vidal2003}. As the interaction parameter $\theta$ moves away from the parameter region of the TL liquid phase, the entropy becomes relatively smaller in Fig. \ref{Fig3}. In Fig. \ref{Fig3}, it should be noted that at $\theta = \pm \pi/2$ and for $-\pi < \theta < -\theta_{c2}$ and $\theta_{c2} < \theta < \pi$, the bipartite entanglement entropy becomes zero, which implies that the ground state becomes a product state. Consequently, our iMPS results show the distinct diverging behavior of the bipartite entanglement entropy for $-\ \theta_{c1}(\infty) < \theta < 0$ and $0 < \theta < \theta_{c1}(\infty)$ in Fig. \ref{Fig4} (a) and otherwise, the saturation behaviors of the bipartite entanglement entropy in Fig. \ref{Fig4} (b). Correspondingly the biquadratic spin-$1$ XY chain with the rhombic anisotropy in Eq. (\ref{ham1}) has a massless phase for $-\ \theta_{c1}(\infty) < \theta < 0$ and $0 < \theta < \theta_{c1}(\infty)$ and otherwise, a massive phase with a mass gap \cite{Vidal2003}. Hence, at $\theta = \pm\, \theta_{c1}$, a gapless to gapped quantum phase transition occurs with the central charge $c \simeq 1$.

\subsubsection{Berezinskii-Kosterlitz-Thouless (BKT)-type quantum phase transition}
 However, as shown in Figs. \ref{Fig1} and \ref{Fig2} (b), the ground state energy and the derivatives of the ground state energy up to the second order are continuous and exhibit no nonanalyticity near the critical points $\theta = \theta_{c1}$ of the bipartite entanglement entropy in Fig. \ref{Fig3}. As a typical example of a gapless to gapped phase transition in one-dimensional spin chain systems, the BKT transition in the one-dimensional spin-$1/2$ XXZ chain model occurs at the critical anisotropy $\Delta = 1$. It has been demonstrated in the spin-$1/2$ XXZ chain model in Refs. \cite{Yang1,Yang2} that the ground state energy and all of the derivatives with respect to the anisotropy $\Delta$ are continuous at the critical point $\Delta=1$ and thus, that is, the transition is an infinite-order quantum phase transition. Hence, in the Hamiltonian (\ref{ham1}), occurring the gapless to gapped phase transition can explain why the derivatives of the ground state energy are continuous and exhibit no nonanalyticity near the critical points $\theta = \theta_{c1}$ in Fige. \ref{Fig1} and \ref{Fig2} (b) because the BKT transition is an infinite-order quantum phase transition, which means the derivatives of the ground state energy does not give any meaningful signal for quantum phase transitions. Similar BKT-type quantum phase transitions were also shown to be detected by using the bipartite entanglement entropy for the quantum phase transition between the gapless TL liquid phase to the gapped dimer phase in the spin-$1/2$ $J_1$-$J_2$ Heisenberg chain \cite{Li2016}, for the transition between the critical XY phase and the antiferromagnetic phase in the spin-$1/2$ XXZ chain model \cite{Choi17}, for the transition between the critical XY phase and the Haldane phase in the spin-$1$ XXZ chain model \cite{Su12,Liu2014,Choi17} and for the massive to massless phase transition in the antiferromagnetic three-state quantum chiral clock model \cite{Dai17}. Accordingly, the gapless to gapped phase transition at $\theta=\pm\, \theta_{c1}$ in the biquadratic spin-$1$ XY chain with the rhombic single-ion anisotropy can then be called BKT-type quantum phase transition.

\subsection{The critical point $\theta =0$ of Tomonaga-Luttinger liquid to liquid quantum phase transition}
 At $\theta = 0$, an inconsistent change of the bipartite entanglement entropy should be noted as the truncation dimension $\chi$ increases from $\chi=30$ to $\chi=150$ in Fig. \ref{Fig3}(a). Specifically, the $S(\chi)$ seems to change without any consistency with increasing of the truncation dimension $\chi$.  Actually, the inconsistent behavior of the amplitude of the $S(\chi)$ is originated from highly degenerate ground states having different entanglement entropies. In other words, for a given truncation dimension $\chi$, the iTBED algorithm \cite{Vidal07} brings a randomly chosen different initial state to a different iMPS ground state with the same ground state energy.  In fact, it has been demonstrated how the iMPS degenerate ground states $|\psi^{(n)}_G\rangle$ can be obtained by using the quantum fidelity in Ref. \cite{Su13}. In this Subsection, we will show that the ground state is highly degenerate at $\theta =0$ and discuss the properties of the ground state.

 \begin{figure}
 \includegraphics[width=0.4\textwidth]{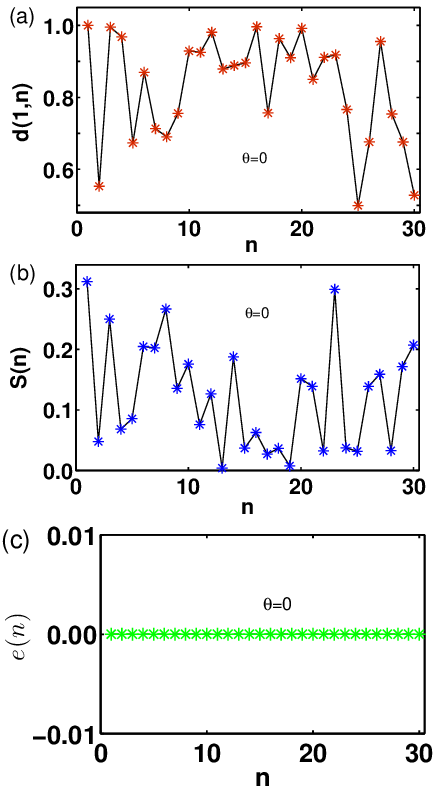}
 \caption{(color online) (a) Quantum fidelity per lattice site $d(1,n)$, (b) bipartite entanglement entropy $S(n)$, and (c) ground state energy $e(n)$ for the iMPS ground states as a function of $n$ at the Tomonaga-Luttinger liquid to liquid transition point $\theta=0$ for the truncation dimension $\chi=30$.  Here, $n=30$ is the number of the random initial state trials.}
 \label{Fig5}
 \end{figure}
 Following Ref. \cite{Su13}, to confirm whether the system has a degenerate ground state for a given parameter $\theta$, one can define a quantum fidelity $F(|\phi\rangle,|\psi^{(n)}_G\rangle) = |\langle\phi | \psi^{(n)}_G\rangle|$ between the ground state $|\psi^{(n)}_G\rangle$ and a chosen reference state $|\phi\rangle$, where $|\psi^{(n)}_G \rangle$ is the iMPS ground state with the $n$th randomly chosen initial state in the iMPS algorithm. The quantum fidelity $F$ asymptotically scales as $F(|\phi\rangle,|\psi^{(n)}\rangle) \sim d^{L}$, where $L$ is the size of the one-dimensional lattice and  the quantum fidelity per lattice site (FLS) $d$ can be defined as
\begin{equation}
  \ln d(|\phi\rangle, |\psi^{(n)}\rangle)
  = \lim_{L \rightarrow \infty}
      \frac{1}{L} \ln F(|\phi\rangle,|\psi^{(n)}\rangle).
  \label{fidelity}
\end{equation}
 The FLS ranges as $0 \leq d(|\phi\rangle, |\psi^{(n)}\rangle) \leq 1$.  Here, we will choose the reference state $|\phi\rangle$ as the ground state $|\psi^{(1)}_G\rangle$ obtained from the first randomly chosen initial state, i.e., $|\phi\rangle = |\psi^{(1)}\rangle$. In this case, the quantum fidelity becomes $F(|\psi^{(1)}_G\rangle, |\psi^{(n)}_G\rangle) = |\langle \psi^{(1)}_G|\psi^{(n)}_G\rangle|$ with $F(|\psi^{(1)}_G\rangle, |\psi^{(1)}_G\rangle)=1$. Thus $d(1,1) = 1 = d(n,n)$.
 Within the iMPS approach, the largest eigenvalue of the transfer matrix corresponds to the FLS. In fact, the FLS is well defined in the thermodynamic limit, even if $F$ becomes trivially zero.

 In order to demonstrate explicitly highly degenerate ground states at $\theta=0$, we calculate 30 ground states from randomly chosen $30$ initial states with the iTBED algorithm for $\chi=30$. In Fig. \ref{Fig5} (a), we plot the FLS as a function of the number of random initial state trials $n$ at $\theta=0$. One can find that the FLSs seem to be different one another but all the ground states have the same energy per site, i.e., $e(n)=0$, as shown in Fig. \ref{Fig5}(c). Obviously, if the system has only one ground state, the FLS also has only one value, i.e., $d(1,n)=1$ independent of the randomly chosen initial state in the iMPS calculation. Accordingly, Fig. \ref{Fig5} (a) shows that since the FLSs have different values one another, there are 30 degenerate ground states. Figure \ref{Fig5} (b) reveals that the bipartite entanglement entropies of the 30 degenerate ground states have various different values, which implies that the degenerate ground states have a very different structure of wave functions one another, although they have the same energy. Accordingly, there are highly degenerate ground states with different entanglement structures at $\theta = 0$.

 The highly degenerate ground states have been shown analytically in the finite-size spin lattice $L$ with the biquadratic spin-$1$ XYZ model $H = \sum_i (J_x S^x_i S^x_{i+1} + J_y S^y_i S^y_{i+1} + J_z S^z_i S^z_{i+1})^2$ under a frustration-free condition \cite{Watanabe24,Saito24,Saito} and the ground state are $L+1$ degenerate in the perspective of a spontaneous breaking of global $U(1)$ symmetry in Refs. \cite{Watanabe24,Shi23-1,Shi23-2}. Thus, an infinite lattice system would have such an infinite degenerate ground states. Moreover, the spin-block entanglement entropy has been numerically shown to be logarithmically divergent with the spin block size $n$  in the thermodynamic limit $L \rightarrow \infty$ as $ S(n) \propto \frac{1}{2} \log_2 n$  \cite{Shi23-2}. The ferromagnetic states with highly degenerate ground states were studied to show a similar entanglement entropy and scaling behaviors \cite{Popkov,Popkov05,Castro}. As for a comparison, let us recall the fact that the central charges in the TLL phases are $c=1$ in Sec. \ref{Luttinger} because the conformal field theory \cite{Affleck91,Holzhey94,Calabrese04} gives such a logarithmically divergent spin block entropy with size of spin block as $S(n) \propto \frac{c}{3} \log_2 n$ for critical systems. In contrast to the transition point $\theta=0$, the spin-block entanglement entropy in the TL liquid phases is to be  $S(n) \propto \frac{1}{3} \log_2 n$.

 Consequently, the biquadratic spin-$1$ XY chain at $\theta=0$ is not in the TL phase and thus undergoes a quantum phase transition across the interaction parameter angle $\theta=0$.
 As were shown in Figs. \ref{Fig1} and \ref{Fig2} (a), the ground state energy and the derivatives of the ground state energy up to the second order are continuous, and a cusp behavior of the second-order derivative seems to happen but the nonanalyticity cannot be confirmed without an explicitly reliable third-order derivative due to the limitation of numerical calculation in this study. This result may suggest that the TL liquid to liquid transition is a third-order quantum phase transition but could also be a higher-order quantum phase transition.

\subsection{Product states}
\label{product}
 In Fig. \ref{Fig3}(a), the bipartite entanglement entropy captures another interesting feature of the ground state, i.e., the zero entanglements at $\theta = \pm\, \pi/2$ and for  $\theta_{c2} < \theta \leq \pi$ and $-\pi \leq \theta < -\theta_{c2}$.  At $\theta =\pm\, \pi/2$, the ground state has the local spin states being the lowest-energy state of the rhombic single-ion anisotropy term, $[(S^x_i)^2 - (S^y_i)^2]$ ($[(S^y_i)^2 - (S^x_i)^2]$) and thus the ground state becomes a uniaxial spin state and the product state of $|S^x_i=0\rangle$ ($|S^y_i=0\rangle$). Such single-site product states give the zero entanglement entropy for the bipartite systems.

 Similarly, at $\theta = \pm \pi$, the ground state with the lowest energy state of the biquadratic term, $-(S_{i}^{x}S_{i+1}^{x}+S_{i}^{y}S_{i+1}^{y})^2$, is in a product state of $|S^z_i=0\rangle$. Thus the local spin fluctuates only in the $xy$ plane, with no change in the $z$ axis. As is found in Fig. \ref{Fig3}(a), the local spin state $|S^z_i=0\rangle$ expands to the transition point $\theta=\pm \theta_{c2} = \pm 0.78\pi$. This implies that the robust local spin fluctuation in the $xy$ plane for the phase protect the ground state structure in the product state of $|S^z_i=0\rangle$ until the rhombic single-ion anisotropy becomes overwhelming the ferromagnetic biquadratic interaction to induce the abrupt change of the spin fluctuation at the transition points $\theta=\pm \theta_{c2}$.  For the phase, the ground state keeps the product state of $|S^z_i=0\rangle$ and thus the bipartite entanglement entropy becomes zero.

\begin{figure}
\includegraphics[width=0.4\textwidth]{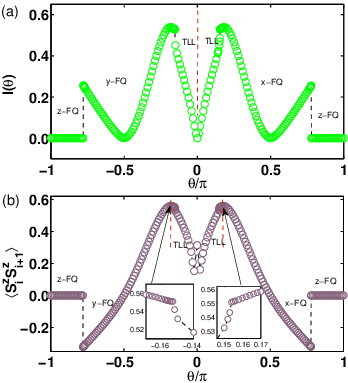}
 \caption{(color online) (a) Mutual information $I$ and (b) spin correlation $\langle S^z_i S^z_{i+1}\rangle$ between the nearest two spins as a function of $\theta$ for $\chi=150$.}
  \label{Fig6}
\end{figure}
\section{Mutual information and spin-spin correlations}
 \label{section5}
 The characteristic divergent or convergent behaviors of the entanglement entropy enable us to distinguish the massless (critical) phases with the central charge $c \simeq 1$ or massive (noncritical) phases, respectively, in the biquadratic spin-$1$ XY chain with the rhombic single-ion anisotropy. Characteristic behaviors of correlations between subsystems of quantum many-body states in each phase provide a deeper understanding of the phases. In our case, the gapless TL liquid phases can also be characterized by a power-law behavior of correlation functions. In this section, we will thus consider the spin-spin correlation as a traditional measure and the quantum mutual information as a quantum information theoretical measure, respectively. In contrast to the conventional spin-spin correlation, the mutual information is used to measure the total amount of correlations, including both quantum and classical correlations, between subsystems. By using the reduced density matrix of the composite state $\varrho_{i\, \cup j}$ for two sites $i$ and $j$ obtained from the iMPS ground state wave function $|\psi_G\rangle$, the spin-spin correlation $C(r)$ and the mutual information $I(r)$ can be defined respectively as
\begin{subequations}
\begin{eqnarray}
 C(|i-j|) &=& \langle \, S_i^z\, S_j^z\, \rangle
 \label{Eta} \\
   I(i:j) &=&  S(i)+S(j)-S(i\cup j),
 \label{Mutual}
\end{eqnarray}
\end{subequations}
 where $S(\alpha) = \mathrm{Tr}\, \varrho_\alpha \log_2 \varrho_\alpha$ is the von Neumann entropy for the lattice sites $\alpha \in \{i, j, i \cup j\}$ and the lattice distance is $r = |i-j|$. Here, $\langle \mathscr{O} \rangle$ stands for the expectation value of an observable operator $\mathscr{O}$ with respect to the ground state wave function $|\psi_G\rangle$.

\begin{figure}
 \includegraphics[width=0.4\textwidth]{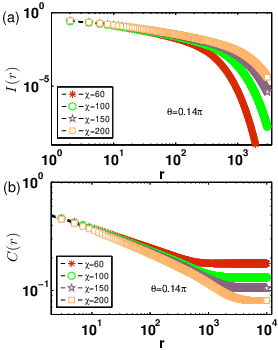}
 \caption{(color online) (a) Mutual information $I(r)$ between two spins $i$ and $j$ with the separation $r=|i-j|$ and (b) spin-spin correlation $C(r)$ of the two spins as a function of the lattice distance $r=|i-j|$ for $\theta=0.14\pi$.
 }
  \label{Fig7}
  \end{figure}
 \subsubsection{Mutual information and spin-spin correlations of the nearest neighbor two spins: Quantum phase transitions}
 Once one obtains the iMPS ground state for the Hamiltonian of Eq. (\ref{ham1}), the quantum mutual information and the correlation can be calculated. In our case, the adjacent two spins are considered. In Fig. \ref{Fig6}, (a) the mutual information $I(\theta)$ and (b) the spin-spin correlation $\langle S^z_iS^z_{i+1}\rangle$ are displayed as a function of $\theta$. The mutual information and the spin correlation exhibit very similar behaviors each other. At $\theta=0$, there is a downward-facing cusp in both the mutual information and the spin correlation. The cusp will be a discontinuity of the first-order derivative of the mutual information and the spin correlation. At $\theta=\pm\, \theta_{c1}$, the kinks appear in the spin correlation and the abrupt jumps arise in the mutual information. Both the mutual information and the spin correlation have the noticeable discontinuities at $\theta=\pm\theta_{c2}$. Both the mutual information and the spin-spin correlation become zero for  $\theta_{c2} < \theta \leq \pi$ and $-\pi \leq \theta < -\theta_{c2}$.

\begin{figure}
\includegraphics[width=0.4\textwidth]{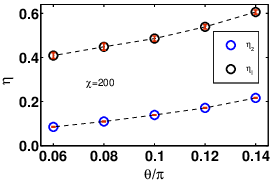}
 \caption{(color online) Critical exponents $\eta_I$ and $\eta_z$ for the mutual information and the spin-sin correlation, respectively, for given parameters $\theta$ and $\chi=200$.
 }
  \label{Fig8}
  \end{figure}
 \subsubsection{Mutual information and spin-spin correlations: Critical exponents in the massless phase}
 As shown in Fig. \ref{Fig6}, the mutual information $I(\theta)$ and spin-spin correlation $\langle S^z_iS^z_{i+1}\rangle$ of the adjacent two spins are symmetric with respect to $\theta=0$. We can focus on the parameter range of positive $\theta > 0$.
 In Figs.~\ref{Fig7} (a) and (b), we plot the mutual information $I(r)$ and the spin-spin correlation $C(r)$ as a function of the lattice distance $r=|i-j|$ at $\theta=0.14\pi$ in the massless phase. Figure \ref{Fig7} (a) shows that the mutual information reveals a power-law decay region for the various truncation dimensions $\chi=60$, $100$, $150$ and $200$. The power-law decay region of the mutual information becomes wider from a few hundreds to a few thousands of the lattice distance as the truncation dimension $\chi$ increases. This tendency of the mutual information implies that the power-law decay range may approach an infinite lattice distance in the thermodynamic limit $\chi \rightarrow \infty$. While the spin-spin correlation exhibits a power-law decay to its saturation value depending on the given truncation dimension $\chi$. Similar to the mutual information, the power-law decay region of the spin-spin correlation becomes wider from a few hundreds to a few thousands of the lattice distance and the saturation value decreases as the truncation dimension $\chi$ increases. In the thermodynamic limit $\chi \rightarrow \infty$, the power-law decay region may approaches an infinite lattice distance and accordingly, the saturation value approaches zero.

 To estimate the critical exponent $\eta_I$  ($\eta_z$) of the mutual information (spin-spin correlation) in the thermodynamic limit, we consider the algebraic decaying part of the mutual information (spin-spin correlation) in Figs. \ref{Fig7}(a) (\ref{Fig7}(b)). The numerical fitting for the power-law decay parts of the mutual information and the spin-spin correlation are performed with the fitting functions $I(r) = a_0 r ^{-\eta_I}$ and $C(r) = b_0 r ^{-\eta_z}$ with the fitting constants $a_0$ and $b_0$. The fitting results of the mutual informations can be summarized as
 (i) $a_0=-0.64(8)$ and $\eta_I= 0.71(2)$ for $\chi=60$,
 (ii) $a_0=-0.69(6)$ and $\eta_I=0.65(1) $ for $\chi=100$,
 (iii) $a_0=-0.72(5)$ and $\eta_I=0.61(1) $ for $\chi=150$,
 and (iv) $a_0=-0.67(4)$ and $\eta_I=0.606(9) $ for $\chi=200$.
 Also, the fitting coefficients of the spin-spin correlations are given as
 (i) $a_0=-0.5291(7)$ and $\eta_z=0.1877(2)$ for $\chi=60$,
 (ii) $a_0=-0.521(3)$ and $\eta_z=0.2029(7)$ for $\chi=100$,
 (iii) $a_0=-0.517(4)$ and $\eta_z=0.209(1)$ for $\chi=150$,
 and (iv) $a_0=-0.513(5)$ and $\eta_z=0.216(1)$ for $\chi=200$.

 In the TL liquid phase, we have performed similar calculations to obtain the exponents of the mutual information and the spin-spin correlations at the four points $\theta=0.06\pi$, $0.08\pi$, $0.1\pi$ and $0.12\pi$. We plot the critical exponents $\eta_I$ and $\eta_z$ for the mutual information and the spin-sin correlation at the five points including $\theta = 0.14 \pi$, respectively, as a function of $\theta$ for $\chi=200$ in Fig. \ref{Fig8}. The numerical estimate $\eta_I$ and $\eta_z$ are given in Table~\ref{table2}. It should be noted that the exponents $\eta_I$ and $\eta_z$ depend on the interaction parameter $\theta$. Figure \ref{Fig8} and Table~\ref{table2} also show that the $\eta_I$ and $\eta_z$ decrease as the interaction parameter $\theta$ decreases. When $\theta$ approaches $\theta=0$, the exponents $\eta_I$ and $\eta_z$ seem to approach zero. Consequently, in the TLL phase of the biquadratic spin-$1$ XY chain with the rhombic single-ion anisotropy, the $\eta_z$ of spin-spin correlation is smaller than the $\eta_I$ of mutual information for a given interaction parameter $\theta$, i.e., $\eta_z < \eta_I$. Similar relationship between the exponents $\eta_z < \eta_I$ was found in the critical line and the critical phase of the spin-$1/2$ XY chain \cite{Dai18}.

\begin{table}
\renewcommand\arraystretch{2}
\caption{Estimate critical exponents $\eta_I$ and $\eta_z$ for the mutual information and the spin-spin correlation in the Tomonaga-Luttinger liquid phases with $\chi=200$.}
\begin{tabular}{cccccccc}
\hline\hline
      \begin{minipage}{1.6cm} $\theta$ \end{minipage}
      &\begin{minipage}{1.1cm}$0.06\pi$ \end{minipage} &
      \begin{minipage}{1.1cm} $0.08\pi$ \end{minipage} &
      \begin{minipage}{1.1cm} $0.1\pi$ \end{minipage} &
      \begin{minipage}{1.1cm} $0.12\pi$ \end{minipage} &
      \begin{minipage}{1.1cm} $0.14\pi$ \end{minipage} \\
\hline
 \begin{minipage}{1.7cm}$\eta_I $  \end{minipage}
 & 0.41(1) & 0.45(2)  & 0.49(1)  & 0.539(9)  & 0.606(9) \\
\hline
 \begin{minipage}{1.7cm}$\eta_z $  \end{minipage}
 &  0.085(1) & 0.109(2)  & 0.139(1)  & 0.171(1)  & 0.216(1) \\
\hline\hline
\end{tabular}
\label{table2}
\end{table}
 In contrast to the gapless TL liquid phases, the mutual information $I(r)$ and the spin-spin correlation exponentially decay to zero in the massive phases $-\pi < \theta < -\theta_{c1}$ and $\theta_{c1} < \theta < \pi$ (not presented here). As a result, the mutual information $I(r)$ and the spin-spin correlation $C(r)$ reveal the characteristic behavior, i.e., algebraic decay to zero with an exponent varying continuously with the interaction parameter ratio $\theta$ for the gapless TL liquid phases (massless phases) $-\theta_{c1} < \theta < 0$ and $0 < \theta < \theta_{c1}$ and exponentially decay to zero with correlation length becoming shorter away from the critical point $\theta_{c1}$ for the massive phases $-\pi < \theta < -\theta_{c1}$ and $\theta_{c1} < \theta < \pi$. Across $\theta = \pm \, \theta_{c1}$, there occur the BKT-type transitions from the massive phases to the TL Liquid phases.

\begin{figure}
\includegraphics[width=0.4\textwidth]{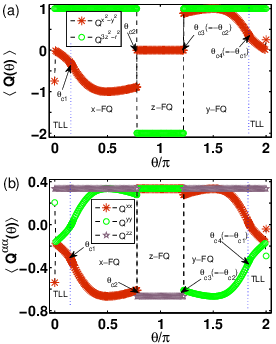}
 \caption{(color online)
 (a) $\langle Q^{x^2-y^2}_i\rangle$ and $\langle Q^{3z^2-r^2}_i \rangle$ and (b) $\langle Q^{\alpha\alpha}_i\rangle$ as a function of $\theta$ with $\alpha \in \{x,y,z\}$ and the truncation dimension $\chi=150$.}
  \label{Fig9}
  \end{figure}
\section{Spin quadrupole moments and Tomonaga-Luttinger liquid state}
 \label{section6}
 In fact, the ground state has no magnetization, i.e., $\langle S^\alpha_i \rangle =0$ except for $\theta =0$. As is shown in Fig. \ref{Fig3} (a), for $-\theta_{c1} < \theta < 0$ and $ 0 < \theta < \theta_{c1}$, there are the two gapless TL liquid (critical) phases in which any local magnetic order vanishes in the thermodynamic limit. Thus, except for the gapless TL liquid regions, a nonmagnetic phase emerges. Also, as is discussed in Sec. \ref{product}, the nonmagnetic phases can be characterized by a characteristic feature of three different product states at $\theta = \pm \pi/2$ and for $\theta_{c2} < \theta \leq \pi$ and $-\pi \leq \theta < -\theta_{c2}$.  A characterization of spin nematic phase with no net magnetization can be performed by discussing behaviors of spin quadrupole moments and their orderings. In this section, let us study and discuss the quadrupole moments in connection with the critical and characteristic behaviors of the bipartite entanglement entropy and the mutual information.

 In order to measure spin quadrupole moments, the quadrupole moment tensor is defined in terms of spin operators at site $i$ as
\begin{equation}
  Q^{\alpha\beta}_i=\frac{1}{2}\left(S^{\alpha}_iS^{\beta}_i+S^{\beta}_iS^{\alpha}_i\right)-\frac{1}{3} \mathbf{S}^2_i\delta^{\alpha\beta}_i,
 \label{Quploe}
\end{equation}
 where $\alpha, \beta \in (x, y, z)$ and $\delta^{\alpha\beta}_i$ is the Kronecker delta at site $i$. Equation \ref{Quploe} shows that the spin quadrupole tensor is symmetric and traceless rank-$2$ tensor, which implies that $Q^{\alpha\beta}_i = Q^{\beta\alpha}_i$ and $\sum_\alpha Q^{\alpha\alpha}=0$. The tracelessness of the quadrupole moment tensor gives the two independent components, i.e., $Q^{x^2-y^2}_i=Q^{xx}_i-Q^{yy}_i$ and $Q^{3z^2-r^2}=3Q^{zz}_i$. Together with $Q^{x^2-y^2}_i$ and $Q^{3z^2-r^2}$, the off-diagonal components $Q^{xy}_i$, $Q^{yz}_i$, and $Q^{zx}_i$ can reveal a characteristic feature of nonmagnetic states.

\subsection{Diagonal quadrupole moments and ferroquadrupole phases}
 Figure \ref{Fig9} (a) displays the two independent diagonal quadrupole orders $\langle Q^{x^2-y^2}_i\rangle$ and $\langle Q^{3z^2-r^2}_i\rangle$ as a function of $\theta$ for the truncation dimension $\chi=150$. Here, the $\theta$ ranges from $\theta=0$ to $\theta=2\pi$. At $\theta= \theta_{c2}$ and $\theta = \theta_{c3} = 2\pi-\theta_{c2}$, the $\langle Q^{x^2-y^2}_i\rangle$ and $\langle Q^{3z^2-r^2}_i\rangle$ change abruptly. These singularities are consistent with those in the ground state energy in Figs. \ref{Fig1} and \ref{Fig2} (c), the bipartite entanglement entropy in Fig. \ref{Fig3} (a), the mutual information and the spin correlation in Fig. \ref{Fig6}. The two discontinuities of the $\langle Q^{x^2-y^2}_i\rangle$ and $\langle Q^{3z^2-r^2}_i\rangle$ divide the parameter range into the three regions. The characteristic features of the $\langle Q^{x^2-y^2}_i\rangle$ and $\langle Q^{3z^2-r^2}_i\rangle$ identify the three spin nematic phases \cite{Mao} as (i) the $x$-FQ phase with  $\langle Q^{x^2-y^2}_i\rangle < 0$ and $\langle Q^{3z^2-r^2}_i\rangle = 1$ for $0 < \theta < \theta_{c2}$, (ii) the $z$-FQ phase with $\langle Q^{x^2-y^2}_i\rangle = 0$ and $\langle Q^{3z^2-r^2}_i\rangle =-2$ for $\theta_{c2} < \theta < \theta_{c3}$, and (iii) the $y$-FQ phase with $\langle Q^{x^2-y^2}_i\rangle > 0$ and $\langle Q^{3z^2-r^2}_i\rangle = 1$ for $\theta_{c3} < \theta < 2\pi$. As a result, $\langle Q^{x^2-y^2}_i\rangle$ can play a role of quadrupole order parameter separating the three $x/y/z$-FQ phases with the phase boundary $\theta = \theta_{c2}$ and $\theta =\theta_{c3}=2\pi -\theta_{c2}$.  The quadrupole moments $\langle Q^{x^2-y^2}_i\rangle$ capture the first-order quantum phase transitions between the spin nematic FQ phases.

\begin{figure}
 \includegraphics[width=0.38\textwidth]{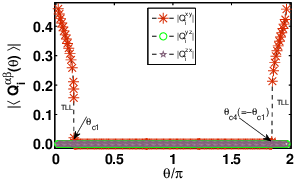}
 \caption{(color online) $|\langle Q_i^{\alpha\beta} \rangle|$ ($\alpha \neq \beta$) at site $i$ as a function of $\theta$ with $\alpha, \beta \in \{x,y,z\}$ for $\chi=150$. Note that at site $i+1$, $\langle Q_{i+1}^{xy} \rangle=-\langle Q_i^{xy} \rangle$ and $\langle Q_{i+1}^{yz} \rangle=\langle Q_i^{zx} \rangle=0$.
 }
 \label{Fig10}
\end{figure}
 However, in contrast to the discontinuities at $\theta= \theta_{c2}$ and $\theta = \theta_{c3}$, the $\langle Q^{x^2-y^2}_i\rangle$ and $\langle Q^{3z^2-r^2}_i\rangle$ are shown to be continuous and exhibit no singularity at $\theta = \theta_{c1}$ and $\theta_{c4}= 2\pi-\theta_{c1}$, independent of the BKT-type transitions between the TL liquid phases and the spin nematic QF phases at $\theta = \theta_{c1}$ and $\theta_{c4}= 2\pi-\theta_{c1}$.
 Moreover, the values of $\langle Q^{x^2-y^2}_i\rangle$ and $\langle Q^{3z^2-r^2}_i\rangle$ are almost unchanged with increasing the truncation dimension and even in the thermodynamics limit. As it should be, the diagonal quadrupole moments satisfy the constraint $\langle Q^{xx}_i\rangle + \langle Q^{yy}_i\rangle + \langle Q^{zz}_i\rangle =0$ or $\langle (S^{x}_i)^2\rangle + \langle (S^{y}_i)^2\rangle + \langle (S^{z}_i)^2\rangle =2$ demonstrated in Fig. \ref{Fig9} (b). This fact implies that the local spins fluctuate under the constraint $\sum_\alpha \langle  (S^{\alpha}_i)^2 \rangle = 0$ in the TL liquid phase. Hence, this shows that the TL liquid state is independent of the local diagonal spin fluctuations $\langle (S^{\alpha}_i)^2 \rangle \neq 0$.

\subsection{Off-diagonal quadrupole moments and Tomonaga-Luttinger liquid phases}
 Interestingly, for roughly biquadratic interaction strength two times stronger than the absolute value of the rhombic single-ion anisotropy, non-zero off-diagonal quadrupole moments emerge. Figure \ref{Fig10} shows clearly that at site $i$, only $\langle Q^{xy}_i \rangle \neq 0$ in the TL liquid phases for $0 < \theta < \theta_{c1}$ and  $\theta_{c4}(=2\pi-\theta_{c1}) < \theta < 2\pi$, otherwise $\langle Q^{yz}_i\rangle =\langle Q^{zx}_i\rangle =0$. Also, at site $i+1$, only $\langle Q^{xy}_{i+1} \rangle = -\langle Q^{xy}_i \rangle$ (not presented here) in the TL liquid phase, otherwise $\langle Q^{yz}_{i+1}\rangle =\langle Q^{zx}_{i+1}\rangle =0$. Nonzero values of $\langle Q^{xy}_i \rangle$ are not sensitive to the truncation dimension $\chi$, nor in the thermodynamic limit. It should be noted that the staggered local spin fluctuations $\langle S^x_i S^y_i\rangle$ occur in the TL liquid phases but do not occur in the nonmagnetic phases. The BKT-type transition points $\theta_{c1}(\chi)$, detected by the nonzero value of $\langle Q^{xy}_i \rangle$, between the TL liquid phases and the nonmagnetic FQ phases are consistent with those of the bipartite entanglement entropy in Fig. \ref{Fig3} (b). Consequently, the staggered local spin fluctuations $\langle S^x_i S^y_i\rangle = - \langle S^x_{i+1} S^y_{i+1}\rangle$ must play a very significant role in the emergence of TL liquid. Thus, the $\langle Q^{xy}_{i/i+1} \rangle$ behave as if the fluctuations act as an order parameter of the TL liquid phase and detect the BKT-type transitions between the TL liquid phase and the spin nematic FQ phases.

\begin{figure}
\includegraphics[width=0.35\textwidth]{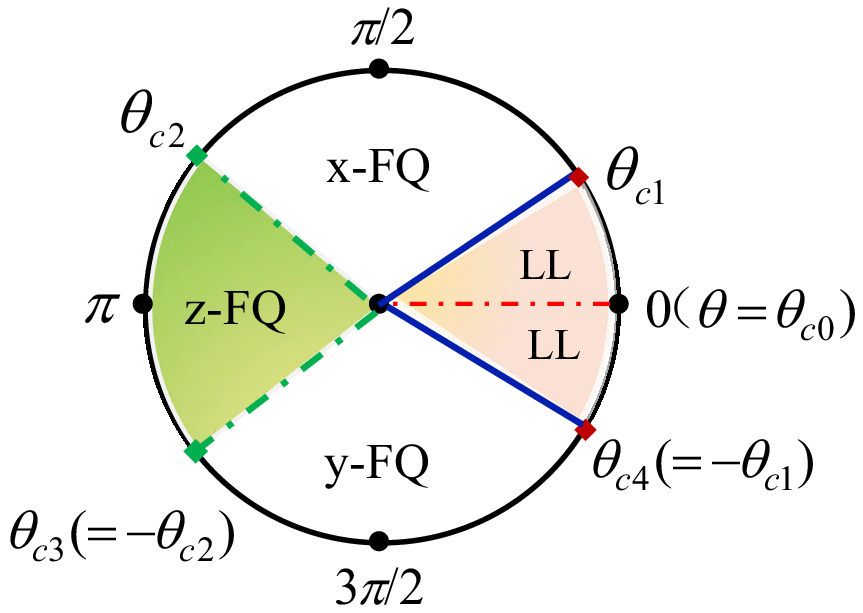}
\caption{(color online)
 Schematic phase diagram of ground state in one-dimensional biquadratic spin-$1$ XY model with rhombic single-ion anisotropy. Quantum phase transitions occur at $\theta_{c0}=0$ and $\theta_{c1/c2} = 2\pi - \theta_{c4/c3}$ in the circular parameter space $\theta$. The $\alpha$-FQ and TL liquid phases stand for the $\alpha$-QF and TL liquid phases, respectively, with $\alpha \in \{x, y, z\}$. The blue solid lines and the red dash-dotted line indicate the infinite-order BKT-type quantum phase transitions and the TL liquid to liquid phase transition, respectively, while the green dash-dotted lines indicate the first-order discontinuous quantum phase transitions between the spin nematic FQ phases. The detailed discussions are in the main texts.}
\label{Fig11}
\end{figure}

\section{Summary}
\label{summary}
 We have investigated the quantum phase transitions in the biquadratic spin-$1$ XY chain with the rhombic single-ion anisotropy numerically by employing the iMPS representation and the iTEBD algorithm. In order to capture the quantum phase transitions and characterize the phases, we have studied systematically with the derivatives of the ground state energy, the bipartite entanglement entropy, the mutual information, and the spin quadrupole moments. As a brief summary, we display the schematic phase diagram of the ground state in Fig. \ref{Fig11}. For the whole parameter range $\theta$, there are the three spin nematic FQ phases and the two gapless TL liquid phases. Mainly three different types of quantum phase transition occur, i.e., (i) the first-order discontinuous quantum phase transitions between the spin nematic FQ phases, (ii) the gapless to gapped quantum phase transitions, i.e., BKT-type quantum phase transitions between the gapless TL liquid phase and the spin nematic FQ phase, and (iii) the continuous quantum phase transition between the two gapless TL liquid phases.

 The ground state energy and the derivatives up to the second order were investigated to detect and classify quantum phase transitions in the spin-$1$ chain. Only the finite discontinuities of the first-order derivatives are shown to capture the first-order discontinuous quantum phase transitions between the spin nematic FQ phases. As it should be, any noticeable singular behavior of the ground state energy and the derivatives has not been revealed for the infinite-order BKT-type quantum phase transitions between the gapless TL liquid phase and the gapped spin nematic FQ phase. The quantum phase transition between the two TL liquid phases has also not been detected up to the second-order derivative of the ground state energy.

 The characteristic divergent or convergent behaviors of the bipartite entanglement entropy are shown to separate the TL liquid phases with the central charge $c \simeq 1$ or the spin nematic FQ phase, respectively. The critical points of the BKT-type quantum phase transitions between the TL liquid phases and the spin nematic FQ phases are estimated using the extrapolation in the thermodynamic limit. We find that the continuous quantum phase transition occur between the two TL liquid phases. Using the quantum fidelity, we show that the ground state is highly degenerate at the TL liquid to liquid transition point. The numerical reliability of higher-order derivatives of the ground state energy prevents us from determining which order quantum phase transition occurs between the two TL liquid phases but cannot eliminate the possibility of a quantum phase transition of higher order than the second order.

 The mutual information and the spin-spin correlation are shown to undergo a power-law decay  with increasing the lattice distance in the TL liquid phases, while they decay exponentially in the spin nematic FQ phases. Moreover, we show that the exponents vary with the interaction parameter $\theta$ of the biquadratic interaction strength and the rhombic single-ion anisotropy. It turns out that the spin-spin correlation exponent $\eta_z$ is smaller than the mutual information one $\eta_I$ , i.e., $\eta_z < \eta_I$ in the TL liquid phases. Accordingly, such a change of characteristic behavior of mutual information and spin-spin correlation indicates an occurrence of the BKT-type quantum phase transition between the TL liquid phase and the spin nematic FQ phase.

 The quadrupole moments are studied in connection with the spin nematic QF phases. The diagonal quardupole moments $\langle Q^{x^2-y^2}_i \rangle$ clearly separate the three different FQ phases, i.e., the $x$-FQ phase with $\langle Q^{x^2-y^2}_i \rangle < 0$, the $z$-FQ phase with $\langle Q^{x^2-y^2}_i \rangle=0$, and the $y$-FQ phase with  $\langle Q^{x^2-y^2}_i \rangle > 0$. Interestingly, the TL liquid phases have a staggered off-diagonal quadrupole moments $\langle Q^{xy}_i\rangle = - \langle Q^{xy}_{i+1} \rangle$. Consequently, in the biquadratic spin-$1$ XY chain with the rhombic single-ion anisotropy, the emergence of the spin fluctuations $\langle S^x_iS^y_i\rangle=-\langle S^x_{i+1}S^y_{i+1}\rangle$  correspond to the advent of the TL liquid phases with the BKT-type quantum phase transitions.

\acknowledgements
 We thank Q.-Q. Shi for helpful discussions. This work was supported in part by Technological Innovation 2030- ``Quantum Communication and Quantum Computer" Major Project, the National Natural Science Foundation of China (Grant No. 11805285) and the Fundamental Research Funds for the Central Universities (Project No.2024CDJXY023, 2019CDXYXDWL0030). Y. H. Su was supported by the Qin Chuangyuan "Scientists + Engineers" Team Project of the Shaanxi Science and Technology Department(Grant No. 2024QCY-KXJ-194)


\end{document}